\documentclass[twocolumn,showpacs,preprintnumbers]{revtex4}
\usepackage{amssymb}
\usepackage{amsmath}
\usepackage{graphicx}
\usepackage{dcolumn}
\usepackage{bm}

\input{tcilatex}

\begin{document}

\preprint{Phys Rev Lett - fluid dynamics}
\title{Length Scales of Acceleration for Locally Isotropic Turbulence}
\author{Reginald J. Hill}
\affiliation{Environmental Technology Laboratory, National Oceanic and Atmospheric
Administration, 325 Broadway, Boulder, Colorado 80305}
\date{\today }

\begin{abstract}
Length scales are determined that govern the behavior at small separations
of the correlations of fluid-particle acceleration, viscous force, and
pressure gradient. \ The length scales and an associated universal constant
are quantified on the basis of published data. \ The length scale governing
pressure spectra at high wave numbers is discussed. \ Fluid-particle
acceleration correlation is governed by two length scales; one arises from
the pressure gradient, the other from the viscous force.
\end{abstract}

\maketitle

\qquad Accelerations in turbulent flow are violent \cite{LaPortaetal} and
are important to many types of studies. \ Experiments in which particles are
tracked in turbulence have found accelerations as large as 1500 times that
of gravity and have quantified the probability density of accelerations to
show that extreme events are likely because the accelerations are highly
intermittent \cite{LaPortaetal}. \ Understanding turbulent accelerations is
essential to many studies, including dispersion and settling of particles in
turbulence \cite{Taylor35}\cite{alternate}, other interactions of bubbles
and particles in turbulence \cite{alternate}, effects on flying insects \cite%
{LaPortaetal}, pollutant transport \cite{Yeung97}, and to the applicability
of Taylor's hypothesis \cite{Heskestad}\cite{Gledzer}. \ Turbulent
accelerations are the object of several studies aimed at solving the mystery
of rain onset from liquid-water clouds; the accelerations affect
collision-coalescence, preferential concentration, and local supersaturation 
\cite{rainman}. \ To advance understanding of turbulent accelerations, the
length scales of acceleration correlations are quantified here, and those
scales are discussed in connection with the scaling of the pressure spectrum
by Gotoh and Fukayama \cite{GotohFukay} and the inertial-range scale found
for fourth-order velocity structure functions by Kerr et al. \cite%
{KerrMeneGot}.

\qquad Universal scaling of the small-scales of turbulence and its analogue
in other areas of physics are discussed in Ref. \cite{SreenAnton97}. \ The
K41 \cite{K41}\ scaling parameters are\ mean energy dissipation rate $%
\varepsilon $ and kinematic viscosity $\nu $. \ Traditional scaling uses K41
scaling of small-scale statistics, and seeks the resultant $R_{\lambda }$\
dependence of the scaled statistic \cite{AntoniaChamSaty81}\cite%
{SreenAnton97}; that $R_{\lambda }$\ dependence is akin to symmetry
breaking. \ Here, $R_{\lambda }\equiv \left\langle u_{1}^{2}\right\rangle
^{1/2}\lambda _{T}/\nu $ is the Reynolds number, where Taylor's \cite%
{Taylor35} length scale is $\lambda _{T}\equiv \left[ \left\langle
u_{1}^{2}\right\rangle /\left\langle \left( \partial u_{1}/\partial
x_{1}\right) ^{2}\right\rangle \right] ^{1/2}$ and $\left\langle
u_{1}^{2}\right\rangle $ is the variance of one component of velocity. \
Since $\left\langle u_{1}^{2}\right\rangle $\ is a property of the
large-scale structure, so is $R_{\lambda }$. \ A radical departure from
tradition is to deduce scaling relationships from relationships amongst
statistics as determined from the equations of motion. \ That method leads,
for example, to the assertion that mean-squared pressure-gradient scales
with an integral of the fourth-order velocity structure function \cite%
{Hill2002}; for such deduced scaling, there is no residual dependence on
large-scale parameters \cite{Hill2002}. \ The deductive method is extended
herein to the length scales of acceleration. \ To quantify the deduced
length scales, it is necessary here as in Ref. \cite{Hill2002} to use
existing turbulence phenomenology; that causes both $\varepsilon $\ and $%
R_{\lambda }$ to appear herein. \ However, the deductive theory does not
explicitly contain K41 scales or $R_{\lambda }$.

\qquad The fluid-particle acceleration $a_{i}$ is related to the
acceleration caused by the viscous force, $\nu \nabla _{\mathbf{x}}^{2}u_{i}$%
, and that caused by the pressure gradient, $-\partial _{x_{i}}p$, by the
Navier Stokes equation: $a_{i}\equiv Du_{i}/Dt=-\partial _{x_{i}}p+\nu
\partial _{x_{n}}\partial _{x_{n}}u_{i}$, where $u_{i}$ is the velocity,$\ p$
is pressure divided by fluid density; density is constant. \ Here, $\nabla _{%
\mathbf{x}}^{2}\equiv \partial _{x_{n}}\partial _{x_{n}}$ is the Laplacian
operator, $\partial $ denotes differentiation with respect to the subscript
variable, summation is implied by repeated Roman indices, \ Consider two
points of measurements, $\mathbf{x}$ and $\mathbf{x}^{\prime }\equiv \mathbf{%
x+r}$; let $r\equiv \left\vert \mathbf{r}\right\vert $, and let prime denote
evaluation at $\mathbf{x}^{\prime }$; e.g., $p^{\prime }\equiv p\left( 
\mathbf{x}^{\prime },t\right) $, $u_{i}^{\prime }\equiv u_{i}\left( \mathbf{x%
}^{\prime },t\right) $, etc. \ Two-point differences are denoted by $\Delta
p\equiv p-p^{\prime }$, $\Delta u_{i}\equiv u_{i}-u_{i}^{\prime }$, etc.;
angle brackets, i.e., $\left\langle \circ \right\rangle $, denote an
average.\ \ Let the 1-axis be in the direction of $\mathbf{r}$ such that the
2- and 3-axes are perpendicular to $\mathbf{r}$. \ Taylor's \cite{Taylor35}
length scale $\lambda _{T}$ is the length scale of the parabola that
osculates the velocity correlation at $\mathbf{r}=0$. \ Specifically, for $%
\mathbf{r=}\left( r_{1},0,0\right) $, $\left\langle u_{1}u_{1}^{\prime
}\right\rangle /\left\langle u_{1}^{2}\right\rangle =1-\left( r_{1}/2\lambda
_{T}\right) ^{2}+\cdot \cdot \cdot $. \ We consider the spatial correlation
tensors of fluid-particle acceleration $\left\langle a_{i}a_{j}^{\prime
}\right\rangle $, pressure gradient $\left\langle \partial _{x_{i}}p\partial
_{x_{j}^{\prime }}p^{\prime }\right\rangle $, and viscous force per unit
mass $\left\langle \nu \nabla _{\mathbf{x}}^{2}u_{i}\nu \nabla _{\mathbf{x}%
^{\prime }}^{2}u_{j}^{\prime }\right\rangle $, as well as the pressure
structure function $D_{P}\left( r\right) \equiv \left\langle \Delta
p^{2}\right\rangle $ and spatial spectra. \ We determine the length scales
of those statistics. \ The length scales considered in this paper are, like $%
\lambda _{T}$, the length scale of the parabolas that osculate those spatial
correlations at $\mathbf{r}=0$. \ Unlike $\lambda _{T}$, all the length
scales defined here do not depend on a large-scale parameter like $%
\left\langle u_{1}^{2}\right\rangle $.

\qquad The considered statistics, $\left\langle \partial _{x_{i}}p\partial
_{x_{j}^{\prime }}p^{\prime }\right\rangle $, $D_{P}$, etc., are related to
the fourth-order velocity structure function (i.e.,\ $\mathbf{D}%
_{ijkl}\left( \mathbf{r}\right) \equiv \left\langle \Delta u_{i}\Delta
u_{j}\Delta u_{k}\Delta u_{l}\right\rangle $) in Ref. \cite{HillWilczak};
that theory is based on the Navier Stokes equation, incompressibility, and
local isotropy, without further assumptions. \ The theory \cite%
{HillThoroddsen} also relates $\left\langle \nu \nabla _{\mathbf{x}%
}^{2}u_{i}\nu \nabla _{\mathbf{x}^{\prime }}^{2}u_{i}^{\prime }\right\rangle 
$ to the third-order velocity structure function $\mathbf{D}_{ijk}\left( 
\mathbf{r}\right) \equiv \left\langle \Delta u_{i}\Delta u_{j}\Delta
u_{k}\right\rangle $. \ The theory has been used to calculate $\left\langle
\partial _{x_{i}}p\partial _{x_{j}^{\prime }}p^{\prime }\right\rangle $ and $%
\left\langle \nu \nabla _{\mathbf{x}}^{2}u_{i}\nu \nabla _{\mathbf{x}%
^{\prime }}^{2}u_{i}^{\prime }\right\rangle $\ from $\mathbf{D}_{ijkl}$\ and 
$\mathbf{D}_{ijk}$, respectively, by means of hot-wire anemometry \cite%
{HillThoroddsen} and from direct numerical simulation (DNS) (see Figs. 12,
13 of Ref. \cite{VedulaYeung}), as well as to compare $D_{P}$\ calculated
from $\mathbf{D}_{ijkl}$\ with $D_{P}$\ calculated from DNS pressure fields 
\cite{HillBoratav}. \ The theory also determines the length scales of those
statistics, which is the topic here.

\qquad Taylor series expansion is used in Ref. \cite{HillWilczak} to
formulate the scaling lengths of $\left\langle \partial _{x_{i}}p\partial
_{x_{j}^{\prime }}p^{\prime }\right\rangle $. \ For brevity, denote the
mean-squared pressure gradient by $\chi \equiv \left\langle \partial
_{x_{i}}p\partial _{x_{i}}p\right\rangle \ $(sum on $i$). \ The longitudinal
component of the pressure-gradient correlation is\ $\left\langle \partial
_{x_{1}}p\partial _{x_{1}^{\prime }}p^{\prime }\right\rangle $ (the 1-axis
is in the direction of $\mathbf{r}$). \ Power series expansion of the
theory's \cite{HillWilczak} relationship between $\left\langle \partial
_{x_{1}}p\partial _{x_{1}^{\prime }}p^{\prime }\right\rangle $\ and
components of $\mathbf{D}_{ijkl}$\ gives 
\begin{align}
\left\langle \partial _{x_{1}}p\partial _{x_{1}^{\prime }}p^{\prime
}\right\rangle & =\left( \chi /3\right) \left( 1-r^{2}/2\lambda
_{1}^{2}+\cdot \cdot \cdot \right) ;  \label{osculate} \\
\lambda _{1}& \equiv \left[ \chi /36\left\langle \left( \partial
_{x_{1}}u_{1}\right) ^{4}\right\rangle h_{Q}\right] ^{1/2};  \label{lambdadp}
\end{align}%
\begin{align}
h_{Q}& \equiv 1+\frac{1}{3}\frac{\left\langle \left( \partial
_{x_{1}}u_{2}\right) ^{4}\right\rangle }{\left\langle \left( \partial
_{x_{1}}u_{1}\right) ^{4}\right\rangle }-3\frac{\left\langle \left( \partial
_{x_{1}}u_{1}\right) ^{2}\left( \partial _{x_{1}}u_{2}\right)
^{2}\right\rangle }{\left\langle \left( \partial _{x_{1}}u_{1}\right)
^{4}\right\rangle }  \label{hQflat} \\
& =\frac{7}{16}\left( 1-\frac{\left\langle \omega _{k}\omega
_{k}s_{ij}s_{ij}\right\rangle }{\left\langle \left( s_{ij}s_{ij}\right)
^{2}\right\rangle }+\frac{1}{4}\frac{\left\langle \left( \omega _{k}\omega
_{k}\right) ^{2}\right\rangle }{\left\langle \left( s_{ij}s_{ij}\right)
^{2}\right\rangle }\right) .  \label{hQsiggia}
\end{align}%
Thus, $\lambda _{1}$ is the scaling length for the longitudinal component $%
\left\langle \partial _{x_{1}}p\partial _{x_{1}^{\prime }}p^{\prime
}\right\rangle ,$ analogous with Taylor's scale. \ In Eq. (\ref{hQsiggia}), $%
h_{Q}$ is given in terms of 3 of the 4 fourth-order invariants determined by
Siggia \cite{Siggia81}, where $s_{ij}$ is the rate of strain tensor and $%
\omega _{k}$ is the vorticity. \ The invariant missing from Eq. (\ref%
{hQsiggia}) is the one that Siggia \cite{Siggia81} expected to vary with $%
R_{\lambda }$ differently than the other 3, so there is no conflict with the
expectation expressed in Ref. \cite{HillWilczak} that $h_{Q}$ becomes a
constant at large Reynolds number. \ Also, note that $h_{Q}=\left\langle
\left( \nabla ^{2}p\right) ^{2}\right\rangle /1575\left\langle \left(
s_{ij}s_{ij}\right) ^{2}\right\rangle $. \ The corresponding length scale of
the transverse component of the pressure gradient correlation is $%
3^{1/2}\lambda _{1}$ (Ref. \cite{HillWilczak} Eq. (61)), and that of $%
\left\langle \partial _{x_{i}}p\partial _{x_{i}^{\prime }}p^{\prime
}\right\rangle $ is $\left( 9/5\right) ^{1/2}\lambda _{1}$. \ The analogous
length scale of $D_{p}\left( r\right) $ is $\sqrt{6}\lambda _{1}$ (Ref. \cite%
{HillWilczak}\ Eq.(38), et seqq.). \ The above shows that to cause
pressure-gradient correlations to coincide near the origin of a graph, one
divides them by $\chi $\ and uses $r/\lambda _{1}$\ on the abscissa; for $%
D_{P}\left( r\right) $, $D_{P}\left( r\right) /\chi \lambda _{1}^{2}$ should
be on the ordinate with $r/\lambda _{1}$\ on the abscissa. \ The pressure
spectrum is the sine transform of the pressure-gradient correlation [Ref. 
\cite{HillWilczak}, Eqs. (15), (22a)]. \ Because $\lambda _{1}$\ scales the
pressure-gradient correlation at small $r$, it follows that to cause
pressure spectra to coincide at viscous-range wave numbers, one should
divide the spectra by $\chi \lambda _{1}^{3}$\ and use $k\lambda _{1}$\ on
the abscissa ($k$ is wave number). \ A Reynolds number limitation for this
to be so is discussed near the end.

\qquad To calculate $\lambda _{1}$\ using Eq. (\ref{lambdadp}) we must
evaluate $h_{Q}$\ using Eqs. (\ref{hQflat}) or (\ref{hQsiggia}). \ Siggia's 
\cite{Siggia81} DNS data for $R_{\lambda }\sim 60$\ to $90$\ gives a single
value $h_{Q}\simeq 0.28$. \ The DNS data of Kerr \cite{Kerr85}\ and the
experimental data of Tsinober \textit{et al}. \cite{Tsinoberetal97}, Pearson
and Antonia \cite{PearsonAntonia}, and Zhou and Antonia \cite{ZhouAnt00},\
can be used to calculate $h_{Q}$; the result is shown in Fig. 1. \ As
suggested by the error bars, the data of Zhou and Antonia do not give a
reliable value of $h_{Q}$, but those data are consistent, within
experimental error, with the other data. The data of Pearson and Antonia 
\cite{PearsonAntonia} (see their Fig. 7), for which error bars are not
available, have scatter similar to those of Ref. \cite{ZhouAnt00}.\ \
However, by fitting their data to straight lines over a broad range of $%
R_{\lambda }$, Pearson and Antonia \cite{PearsonAntonia} mitigate random
error to find that for $37<R_{\lambda }<10^{3}$, $\left\langle \left(
\partial _{x_{1}}u_{2}\right) ^{4}\right\rangle /\left\langle \left(
\partial _{x_{1}}u_{1}\right) ^{4}\right\rangle =5.52$ and $\left\langle
\left( \partial _{x_{1}}u_{1}\right) ^{2}\left( \partial
_{x_{1}}u_{2}\right) ^{2}\right\rangle /\left\langle \left( \partial
_{x_{1}}u_{1}\right) ^{4}\right\rangle =0.84$. \ Use of Eq. (\ref{hQflat})
then gives $h_{Q}=0.32$; that value is shown as the solid line extending
over the range $37<R_{\lambda }<10^{3}$ in Fig. 1. \ Note the slight
increase of $h_{Q}$\ with decreasing $R_{\lambda }$ at $R_{\lambda }<20$. \
The data of Refs. \cite{Kerr85}-\cite{PearsonAntonia} are in such good
agreement that $h_{Q}$ is probably independent of $R_{\lambda }$\ for the
range $20<R_{\lambda }<10^{3}$, wherein $h_{Q}\simeq 0.3$, and this value
might persist for $R_{\lambda }\rightarrow \infty $.\FRAME{ftbpFU}{1.222in}{%
0.8735in}{0pt}{\Qcb{The $R_{\protect\lambda }$ dependence of $h_{Q}$: $\ +$%
,\ DNS data \protect\cite{Kerr85}; experimental data: $\triangle $, 
\protect\cite{Tsinoberetal97}; $\square $, \protect\cite{ZhouAnt00} with
error bars; solid line, \protect\cite{PearsonAntonia}.}}{}{hillprlfig1.ps}{%
\special{language "Scientific Word";type "GRAPHIC";maintain-aspect-ratio
TRUE;display "USEDEF";valid_file "F";width 1.222in;height 0.8735in;depth
0pt;original-width 6.9998in;original-height 4.9926in;cropleft "0";croptop
"1";cropright "1";cropbottom "0";filename
'D:/RSI/IDL54/HillprlFig1.ps';file-properties "XNPEU";}}

\qquad Because $\lambda _{1}$\ is essential for scaling, we next quantify it
by means of existing phenomenology. \ Use of existing phenomenology requires
use of the K41 acceleration scale, $\varepsilon ^{3/4}\nu ^{-1/4}$,\ and
introduction of $\eta \equiv \left( \nu ^{3}/\varepsilon \right) ^{1/4}$, $%
\varepsilon =15\nu \left\langle \left( \partial _{x_{1}}u_{1}\right)
^{2}\right\rangle $, and $F\equiv \left\langle \left( \partial
_{x_{1}}u_{1}\right) ^{4}\right\rangle /\left\langle \left( \partial
_{x_{1}}u_{1}\right) ^{2}\right\rangle ^{2}$\ into Eq. (\ref{lambdadp}). \
Doing so reexpresses (\ref{lambdadp}), without approximation, as%
\begin{equation}
\lambda _{1}=\left( 5/2\right) \eta \left( \chi /\varepsilon ^{3/2}\nu
^{-1/2}\right) ^{1/2}\left( Fh_{Q}\right) ^{-1/2}.  \label{general}
\end{equation}%
We emphasize that the theory \cite{HillWilczak}\ does not contain K41
scaling. \ Use of Eq. (\ref{general}) leads to asymptotic expressions for $%
\lambda _{1}$\ for high and low $R_{\lambda }$ as follows. \ On the basis of
empirical data for the form of $D_{1111}\left( r\right) $, an approximation
for $\chi $ is $\chi \simeq 3.1H_{\chi }\varepsilon ^{3/2}\nu ^{-1/2}F^{0.79}
$\ for\ $R_{\lambda }\gtrsim 400$ \cite{Hill2002}. \ This formula produced
quantitative agreement \cite{Hill2002} with the DNS data of Refs. \cite%
{VedulaYeung} and \cite{GotohFukay} when combined with the data for $F$ at
high Reynolds numbers from Ref. \cite{AntoniaChamSaty81}, namely $F\simeq
1.36R_{\lambda }^{0.31}$. \ Recently, the same DNS data as that of Ref. \cite%
{GotohFukay} have been used to give values of $F$ in Ref. \cite{GotohFukNak}%
. \ Those values at $R_{\lambda }\gtrsim 400$\ are consistent with $F\simeq
1.18R_{\lambda }^{0.31}$. \ To use $F\simeq 1.18R_{\lambda }^{0.31}$\
instead of $F\simeq 1.36R_{\lambda }^{0.31}$\ and to maintain quantitative
agreement of $\chi $ with the DNS data of Refs. \cite{VedulaYeung}-\cite%
{GotohFukay}, it is necessary to use the coefficient $3.5$ instead of $3.1$
in $\chi \simeq 3.1H_{\chi }\varepsilon ^{3/2}\nu ^{-1/2}F^{0.79}$. \ That
is, to be consistent with the data of both Refs. \cite{GotohFukNak} and \cite%
{GotohFukay}, we use,\ for $R_{\lambda }\gtrsim 400$,%
\begin{equation}
\chi \simeq 3.5H_{\chi }\varepsilon ^{3/2}\nu ^{-1/2}F^{0.79}\simeq
4.4H_{\chi }\varepsilon ^{3/2}\nu ^{-1/2}R_{\lambda }^{0.25}.  \label{hiRchi}
\end{equation}%
Evidence that $H_{\chi }$\ is a constant for $R_{\lambda }\gtrsim 80$ and
that its value is about $0.65$\ is given in Ref. \cite{Hill2002}. \ For $%
H_{\chi }=0.65$, Eq. (\ref{hiRchi}) agrees quantitatively with the DNS data
in Table 1 of Ref. \cite{GotohFukay} for $R_{\lambda }\geq 387$; also, Eq. (%
\ref{hiRchi}) is a good approximation of the DNS data for $R_{\lambda }$\ as
small as $200$.\ \ Substitute Eq. (\ref{hiRchi}) in Eq. (\ref{general}) to
obtain, for $R_{\lambda }>400$, 
\begin{equation}
\lambda _{1}\simeq 4.7\eta \left( H_{\chi }/F^{0.21}h_{Q}\right)
^{1/2}\simeq 6.8\eta R_{\lambda }^{-0.033}.  \label{highlambda1}
\end{equation}%
For the right-most expression in Eq. (\ref{highlambda1}), $H_{\chi }=0.65$, $%
h_{Q}=0.3$, and $F\simeq 1.18R_{\lambda }^{0.31}$\ were used. \ For the case
of low Reynolds numbers, Ref. \cite{Hill2002} obtains 
\begin{equation}
\chi \simeq 0.11\varepsilon ^{3/2}\nu ^{-1/2}R_{\lambda }\text{ \ for \ }%
R_{\lambda }\lesssim 20,  \label{loRchi}
\end{equation}%
which agrees with the DNS data of Ref. \cite{VedulaYeung}. \ For low
Reynolds numbers, use Eq. (\ref{loRchi}) in Eq. (\ref{lambdadp}) to obtain 
\begin{equation}
\lambda _{1}\simeq 0.83\eta \left( R_{\lambda }/Fh_{Q}\right) ^{1/2}\text{ \
for \ }R_{\lambda }<20.  \label{lolambda1}
\end{equation}%
None of Eqs. (\ref{hiRchi}) to (\ref{lolambda1}) is K41 scaling because of
their $R_{\lambda }$\ dependence. \ Figure 2 shows $\lambda _{1}$\
increasing relative to $\eta $\ as $R_{\lambda }$\ increases from $9$ to
attain a maximum near $R_{\lambda }\approx 150$, beyond which $\lambda _{1}$%
\ gradually decreases relative to $\eta $\ in agreement with Eq. (\ref%
{highlambda1}).\FRAME{ftbpFU}{1.3266in}{0.947in}{0pt}{\Qcb{The $R_{\protect%
\lambda }$ dependence of $\protect\lambda _{1}/\protect\eta $: $\ +$, DNS
data \protect\cite{Kerr85} in Eq. (\protect\ref{lolambda1}); $\ast $, DNS
data \protect\cite{VedulaYeung} in Eq. (\protect\ref{general}); $\triangle $%
, DNS data \protect\cite{GotohFukay},\protect\cite{GotohFukNak} in Eq. (%
\protect\ref{general}); -----, Eq. (\protect\ref{highlambda1}).}}{}{%
lambda1vsr.ps}{\special{language "Scientific Word";type
"GRAPHIC";maintain-aspect-ratio TRUE;display "USEDEF";valid_file "F";width
1.3266in;height 0.947in;depth 0pt;original-width 6.9998in;original-height
4.9926in;cropleft "0";croptop "1";cropright "1";cropbottom "0";filename
'D:/RSI/IDL54/Lambda1vsR.ps';file-properties "XNPEU";}}

\qquad We now consider the spatial correlation of the viscous force because
it it is part of the fluid-particle correlation as well as of intrinsic
interest \cite{HillAlternativ}. \ The spatial correlation of $\nu \nabla _{%
\mathbf{x}}^{2}\mathbf{u}$ is \cite{HillThoroddsen}\cite{HillAlternativ}: 
\begin{eqnarray}
V_{ij}\left( \mathbf{r}\right) &\equiv &\left\langle \nu ^{2}\nabla _{%
\mathbf{x}}^{2}u_{i}\nabla _{\mathbf{x}^{\prime }}^{2}u_{i}^{\prime
}\right\rangle =-\frac{\nu ^{2}}{2}\nabla _{\mathbf{r}}^{2}\nabla _{\mathbf{r%
}}^{2}D_{ij}\left( \mathbf{r}\right)  \label{Vij(r)} \\
&=&-\left( \nu /4\right) \nabla _{\mathbf{r}}^{2}\partial
_{r_{k}}D_{ijk}\left( \mathbf{r}\right) ,  \label{Moninseq}
\end{eqnarray}%
where $\nabla _{\mathbf{r}}^{2}$ is the Laplacian operator in $\mathbf{r}$%
-space. \ Then, for the longitudinal components of $V_{ij}\left( \mathbf{r}%
\right) $, Eq. (\ref{Moninseq}) yields \cite{HillThoroddsen} 
\begin{subequations}
\begin{eqnarray}
V_{11}\left( r\right) &=&\left( \nu /2r^{3}\right) \text{[}D_{111}\left(
r\right) +2D_{122}\left( r\right)  \notag \\
&&-5r\partial _{r}D_{122}\left( r\right) -r^{2}\partial
_{r}^{2}D_{122}\left( r\right) \text{]}.  \label{V11(r)}
\end{eqnarray}%
A similar formula applies to the transverse component $V_{22}\left( r\right) 
$ \cite{HillThoroddsen}. \ Power series expansion gives $D_{111}\left(
r\right) =\left\langle \left( \partial _{x_{1}}u_{1}\right)
^{3}\right\rangle r^{3}\left[ 1-\left( r^{2}/\left( 2\lambda
_{D_{111}}^{2}\right) \right) +\cdot \cdot \cdot \right] $, where 
\end{subequations}
\begin{equation}
\lambda _{D_{111}}=\left[ \frac{-4\left\langle \left( \partial
_{x_{1}}u_{1}\right) ^{3}\right\rangle }{3\left\langle \left( \partial
_{x_{1}}^{2}u_{1}\right) ^{2}\partial _{x_{1}}u_{1}\right\rangle
+2\left\langle \left( \partial _{x_{1}}u_{1}\right) ^{2}\partial
_{x_{1}}^{3}u_{1}\right\rangle }\right] ^{1/2}.  \label{lambdaD111}
\end{equation}%
Then, Eq. (\ref{V11(r)}) yields 
\begin{equation*}
V_{11}\left( r\right) =-\frac{35}{6}\nu \left\langle \left( \partial
_{x_{1}}u_{1}\right) ^{3}\right\rangle \left[ 1-\left( r^{2}/2\lambda
_{V_{11}}^{2}\right) +\cdot \cdot \cdot \right] ,
\end{equation*}%
where $\lambda _{V_{11}}=\lambda _{D_{111}}/\sqrt{18/5}$ is the length scale
of the osculating parabola of $V_{11}\left( r\right) $. $\ $The
corresponding expansions of the transverse component of $V_{ij}\left( 
\mathbf{r}\right) $ and $V_{ii}\left( r\right) $ give length scales $\lambda
_{V_{11}}/\sqrt{2}$ and $\lambda _{V_{11}}/\sqrt{5/3}$, respectively. \
Kolmogorov's equation can relate the velocity-derivative moments in Eq. (\ref%
{lambdaD111}) to other high-order derivative moments; this gives the same
result as would beginning with the relationship of $V_{ij}\left( \mathbf{r}%
\right) $\ to $D_{ij}\left( \mathbf{r}\right) $ in Eq. (\ref{Vij(r)}). \ The
difficulty of observing the initial fall-off of $V_{11}\left( r\right) $,
and hence its length scale $\lambda _{V_{11}}$ is clearly illustrated in
Fig. 3 of Ref. \cite{HillThoroddsen} and Fig. 13 of Ref. \cite{VedulaYeung}
and their discussions; one requires a spatial resolution substantially finer
than $\eta $. \ Such fine resolution data is not yet available. \ The
high-order derivative moments in the denominator of Eq. (\ref{lambdaD111})
likewise require fine spatial resolution; their Reynolds-number dependence
has not been investigated, but it would be surprising if they obey K41
scaling.

\qquad Despite these difficulties, we can obtain the $r$-values at which $%
V_{11}\left( r\right) /V_{11}\left( 0\right) =0.5$, and similarly for the
transverse component,\ from the graphs of those correlation coefficients in
Fig. 13 of Ref. \cite{VedulaYeung}; the $r$-values are $3.5\eta $ and $%
2.5\eta $ for longitudinal and transverse components, respectively. \
Limited to $R_{\lambda }=230$\ of the DNS run in Fig. 13 of Ref. \cite%
{VedulaYeung}, $3.5\eta $ and $2.5\eta $ are the first estimates of these
length scales. \ It is not suggested that these length scales maintain fixed
ratio to $\eta $\ as $R_{\lambda }$\ varies.

\qquad The fluid-particle acceleration correlation is \cite{HillWilczak} 
\begin{equation}
A_{ij}\left( \mathbf{r}\right) \equiv \left\langle a_{i}a_{j}^{\prime
}\right\rangle =\left\langle \partial _{x_{i}}p\partial _{x_{j}^{\prime
}}p^{\prime }\right\rangle +V_{ij}\left( \mathbf{r}\right) .
\label{accelcorr}
\end{equation}%
As a result of the differing length scales of $\left\langle \partial
_{x_{i}}p\partial _{x_{j}^{\prime }}p^{\prime }\right\rangle $\ and $%
V_{ij}\left( \mathbf{r}\right) $, the theory requires that the initial
fall-off of the longitudinal component $A_{11}\left( r\right) $\ be
described by two length scales; the same is true for the transverse
component and for $A_{ii}\left( r\right) $. \ For example, consider the
trace $A_{ii}\left( r\right) $ and $R_{\lambda }=230$. \ From the above
discussion we have $\lambda _{V_{11}}=3.5\eta $ such that $\lambda
_{V_{ii}}=\lambda _{V_{11}}/\sqrt{5/3}=2.7\eta $. \ From Fig. 2 at $%
R_{\lambda }=230$\ we have that the length scale of $\left\langle \partial
_{x_{i}}p\partial _{x_{i}^{\prime }}p^{\prime }\right\rangle $ is $\sqrt{9/5}%
\lambda _{1}=7.6\eta $. \ Also, $V_{ii}\left( 0\right) /\chi =0.015$ at $%
R_{\lambda }=230$ \cite{Hill2002}\cite{VedulaYeung}. \ Thus, $A_{ii}\left(
r\right) $ has an initial rapid but small-amplitude decay (1.5\% of the
total) from $V_{ii}\left( r\right) $, having scale $2.7\eta $, followed by
the larger amplitude and more gradual decay of $\left\langle \partial
_{x_{i}}p\partial _{x_{i}^{\prime }}p^{\prime }\right\rangle $\ with scale $%
7.6\eta $. \ It is not implied that the two length scales maintain fixed
ratio as $R_{\lambda }$\ varies; the example above is for $R_{\lambda }=230$.

\qquad We noted above that to cause pressure spectra to coincide at
viscous-range wave numbers, one should divide the spectra by $\chi \lambda
_{1}^{3}$\ and use $k\lambda _{1}$\ on the abscissa. \ However, the Reynolds
number must be large enough that, at high wave numbers, the pressure
spectrum has negligible contributions from the sine transform of the
pressure-gradient correlation at$\ r$ on the order of the integral scale. \
How large is large enough? \ From the pressure spectra from isotropic DNS
shown in Fig. 5 of Gotoh and Fukayama \cite{GotohFukay}, where $k\eta $\ is
the abscissa, combined with the weak variation of $\lambda _{1}/\eta $ shown
in Fig. 2 for $100<R_{\lambda }<400$, it seems that $R_{\lambda }\gtrsim 200$
is large enough. \ Of course, no such limitation applies to scaling of $%
\left\langle \partial _{x_{i}}p\partial _{x_{j}^{\prime }}p^{\prime
}\right\rangle $ and $D_{p}\left( r\right) $ with the parameters $\chi $\
and $\lambda _{1}$.

\qquad Kerr \textit{et al}. \cite{KerrMeneGot}\textit{\ }identify, by
empirical means, a length scale within the inertial range of fourth-order
velocity structure functions ($\mathbf{D}_{ijkl}\left( \mathbf{r}\right) $)
such that scaling exponents should be determined only from $r$ greater than
that scale. \ In the inertial range, the divergence of $\mathbf{D}%
_{ijkl}\left( \mathbf{r}\right) $\ equals the pressure-gradient
velocity-velocity structure function which is, in turn, related to $%
A_{ij}\left( \mathbf{r}\right) $ (Eqs. 9, 10, A4, A5 of Ref. \cite{hillBor01}%
). \ Taylor series expansion of those relationships shows that $\lambda _{1}$%
\ is a scale of, at least, linear combinations of the nonzero components of $%
\mathbf{D}_{ijkl}\left( \mathbf{r}\right) $. \ From Fig. 2 of Ref. \cite%
{KerrMeneGot}, their empirical scale is about 5 times $\lambda _{1}$; on the
other hand, $\lambda _{1}$ is only the initial roll-off of $\left\langle
\partial _{x_{i}}p\partial _{x_{j}^{\prime }}p^{\prime }\right\rangle $. \
Further work is needed to establish a causal relationship between the
scales, if one exists.

\qquad The theory \cite{HillWilczak} predicts the scaling with $\chi $ that
was found empirically in Refs. \cite{VedulaYeung} and \cite{GotohFukay}. \
Even for large Reynolds numbers, dividing $\left\langle \partial
_{x_{1}}p\partial _{x_{1}^{\prime }}p^{\prime }\right\rangle $ by $\chi $\
and $D_{P}\left( r\right) $ by $\chi \lambda _{1}^{2}$ and similarly for
pressure spectra\ is not K41 scaling. \ The approximation Eq. (\ref{hiRchi})
contains, in addition to K41 scaling parameters, the factor $F$ and thereby
dependence on $R_{\lambda }$. \ The deviation of both $\chi $\ and $F$\ from
K41 scaling is large for large variations of Reynolds number. \ Whether or
not $\lambda _{1}/\eta $ continues the weak downward trend shown in Fig. 2
or becomes constant as $R_{\lambda }\rightarrow \infty $\ is also a topic
for further investigation, as is the relationship of $\eta $\ to the
viscous-force scale $\lambda _{V_{11}}$.

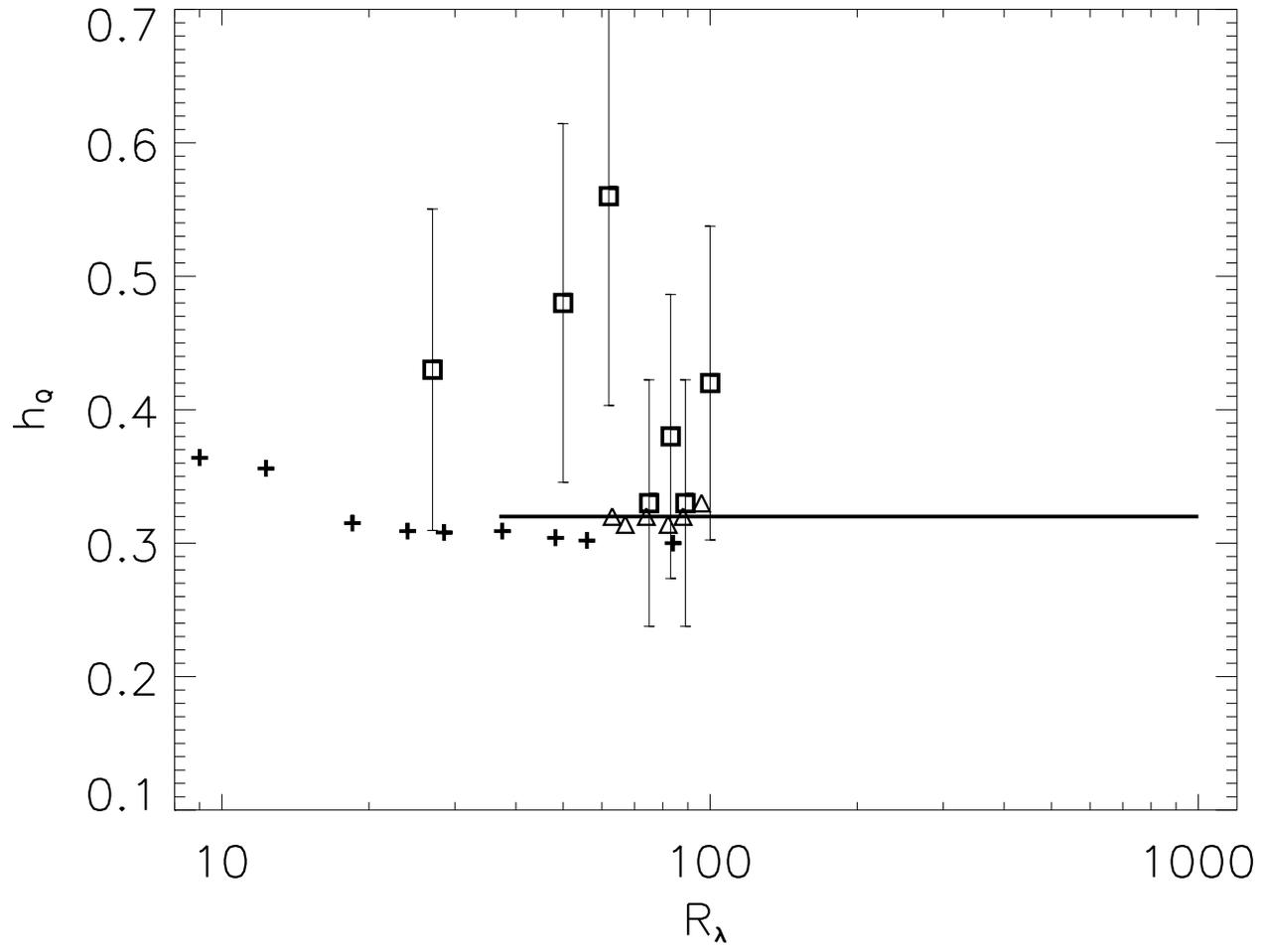
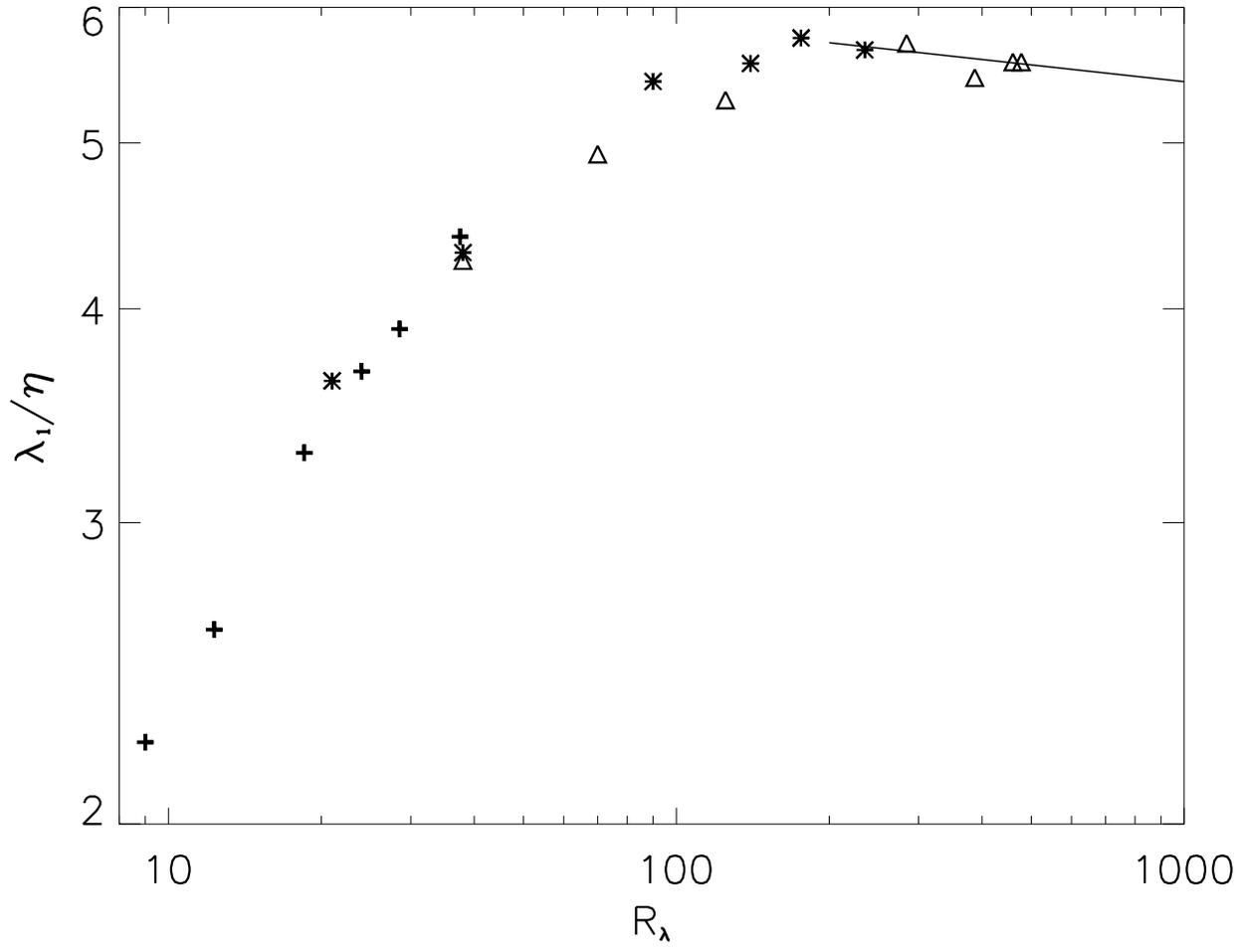
\end{document}